\documentclass[12pt]{article}
\usepackage{graphicx}
\usepackage{array}
\usepackage{multirow}
\usepackage{amsfonts}
\usepackage{amssymb}
\usepackage{mathrsfs}
\usepackage{amsmath}
\usepackage{subfigure}
\makeatletter

\newcommand{\Rmnum}[1]{\expandafter\@slowromancap\romannumeral #1@}
\makeatother

\usepackage[unicode]{hyperref}
\usepackage[numbers,sort&compress]{natbib}

\begin{document}
\renewcommand{\thefootnote}{\fnsymbol{footnote}}
\begin{titlepage}

\vspace{10mm}
\begin{center}
{\Large\bf Charged Lovelock black holes in the presence of dark fluid with a nonlinear equation of state}
\vspace{9mm}
%%{{\large Xiang-Qian Li${}^{1,2,}$\footnote{E-mail address: lixiangqian@tyut.edu.cn}},

{{\large Xiang-Qian Li${}^{1,2,}$\footnote{E-mail address: lixiangqian@tyut.edu.cn}},
{\large Bo Chen${}^{1,2}$},
{\large Li-li Xing${}^{1,2}$}\\

\vspace{6mm}
${}^{1}${\normalsize \em College of Physics and Optoelectronic Engineering,
Taiyuan University of Technology, Taiyuan 030024, China
}

\vspace{3mm}
${}^{2}${\normalsize \em Key Laboratory of Advanced Transducers and Intelligent Control System, Ministry of Education and Shanxi Province, Taiyuan University of Technology, Taiyuan 030024,China}
}

\end{center}

\vspace{10mm}
\centerline{{\bf{Abstract}}}
\vspace{6mm}
\noindent

We consider a model that charged static spherically-symmetric black hole is surrounded by dark fluid with nonlinear equation of state $p_d=-B/\rho_d$. We find that the energy density of the dark fluid can be characterized by two parameters. The derivation of metric solution, as well as the calculation of black hole thermodynamical quantities as functions of horizon radius, are performed. Specially, in $D$-dimensional Einstein gravity and Gauss-Bonnet gravity cases, we plot the metric functions and corresponding thermodynamical quantities, such as mass, Hawking temperature and heat capacity, by varying the values of spacetime dimensions and dark fluid parameters. The effects of the dark fluid parameters on black hole solutions as well as on thermodynamical stability of black holes are discussed.

\vskip 20pt
\noindent
{\bf PACS Number(s)}: 04.20.Cv, 04.50.Gh, 95.35.+d


\end{titlepage}
\newpage
\renewcommand{\thefootnote}{\arabic{footnote}}
\setcounter{footnote}{0}
\setcounter{page}{2}

\section{Introduction}
As well known, Einstein's general relativity has been extensively tested at low energy scale \cite{reviewofGRtests}. However, it is expected that Einstein's theory should be modified at very high energies close to the Planck scale. String theory \cite{stringtheory} and brane cosmology \cite{branecosmology} strongly predict the existence of extra dimensions and therefore generalizing the gravity theory to higher dimensions seems worthwhile. Lovelock \cite{Lovelock} introduced a theory that adds higher curvature terms to the Einstein-Hilbert action in higher dimensions. It is worth to mention that the Lovelock gravity does not contain terms with more than second derivatives of metric, and thus free of ghost \cite{freeghost}. In four dimensions, Lovelock gravity just recovers to the general relativity; while in five dimensions, the quadratic curvature term which is known as the Gauss-Bonnet term in string theory \cite{freeghost,GBterm}, appears in the gravitational action. Some black hole solutions in Lovelock gravity theories were obtained in the literature \cite{BoulwarePRL1985,WheelerNPB1986,MyersPRD1988,CaiPLB2004,Li:2011uk,Cvetic:2016sow,HennigarJHEP2017,HennigarPRL2017}.

At present time there are numerous observational data firmly indicating that the Universe is undergoing a phase of accelerated expansion which may be due to dark energy that is gravitationally self-repulsive. The 2018 release of Planck data about CMB power spectra and CMB lensing, in combination with BAO, show that baryon matter component is no more than $5\%$ for total energy density, and about $95\%$ energy density in the Universe is invisible which includes dark energy and dark matter \cite{Planck2018}. The dominance of the dark sector over the Universe makes the study of black holes surrounded by these mysterious field important. Quintessence is a possible candidate for dark energy, which is characterized by the linear equation of state $p_q=w\rho_q$, where $p_q$ is the pressure, $\rho_q$ is the energy density, and  $-1 < w < -1/3$. Significant attention has been devoted to discussion of static spherically-symmetric black hole solutions surrounded by quintessence matter and their properties \cite{KiselevCQG2003,MaCPL2007,FernandoGRG2012,FengPLA2014,MalakolkalamiASS2015,HussainGRG2015,Ghosh:2016ddh,Ghosh:2017cuq,Toledo:2018pfy}, within which, Refs. \cite{Ghosh:2016ddh,Ghosh:2017cuq,Toledo:2018pfy} paid attention to the framework of Lovelock gravity.

With regard to the Universal dark sector, there exists another possibility that the unknown energy component is a unified dark fluid, i.e. a hybrid of dark matter and dark energy. Among the proposed unified dark fluid models, the Chaplygin gas (CG) \cite{Kamenshchik:2001cp} and its generalized model \cite{Bilic:2001cg,Bento:2002ps} have been widely studied in order to explain the accelerating Universe \cite{Carturan:2002si,Amendola:2003bz,Bean:2003ae}. This dark fluid has the fascinating property that it dynamically behaves as dark matter in the early time and as dark energy in the late time. In order to achieve that, the dark fluid is characterized by using an exotic nonlinear equation of state. In this paper we study the static spherically-symmetric black holes in the presence of electrostatic field and Chaplygin-like dark fluid with the equation of state $p_d=-B/\rho_d$ in the framework of Lovelock gravity, where the subscript `$d$' denotes the dark fluid.

The plan of this paper is as follows. In section \ref{section2}, we give a brief introduction to the Lovelock gravity. In section \ref{section3}, for the dark fluid with a nonlinear equation of state in $D$-dimensional spacetime, we deduce its energy momentum tensor. In section \ref{section4}, we study the static spherically-symmetric Lovelock spacetime in the presence of central electric charge and surrounding dark fluid, further we calculate the thermodynamical quantities corresponding to the new derived black hole solution. Black holes in $D$-dimensional Einstein gravity and Gauss-Bonnet gravity, as two special cases, are also studied in section \ref{section4}. Section \ref{section5} gives the concluding remarks.

We use units which fix the speed of light and the gravitational constant via $8\pi G = c = 1$, and use
the metric signature ($-,\;+,\;+,\; \cdots,\;+$).

\section{Lovelock gravity}
\label{section2}

Lovelock gravity theories are fascinating extensions of general relativity that include
higher curvature interactions. The action of the general Lovelock gravity reads
\begin{equation}
S=S_L+S_M,\label{action}
\end{equation}
where $S_M$ represents the action for the electrostatic field and the dark fluid, since we want to model charged Lovelock black holes surrounded by dark fluid with a nonlinear equation of state in this paper.
The  $S_L$ in $D$ spacetime dimension reads
\begin{equation}
S_L=\frac{1}{2}\int dx^{D}\sqrt{-g}\mathcal{L},
\end{equation}
with Lagrangian $\mathcal{L}$
\begin{equation}
\mathcal{L}=c_0+R+\sum\limits_{k=2}^{m}c_{k}\mathcal{L}_k,
\end{equation}
where $c_0$ relates to the cosmological constant $\Lambda$ by $c_0=-2\Lambda$, $c_{k}$ is a coupling constant, $m=[(D-1)/2]$ denotes the integral part of $(D-1)/2$, $\mathcal{L}_k$ is the Euler density given by
\begin{equation}
\mathcal{L}_k=\frac{1}{2^{k}}\delta _{\alpha _{1}\beta_{1}...
\alpha _{k}\beta _{k}}^{\mu _{1}\nu _{1}...\mu _{k}\nu_{k}}
R^{\alpha _{1}\beta _{1}}{}_{\mu _{1}\nu _{1}}\cdots R^{\alpha _{k}\beta _{k}}{}_{\mu _{k}\nu _{k}}.\label{EulerDensity}
\end{equation}
The $\delta$ symbol above denotes the totally antisymmetric product of $2k$ Kronecker deltas, normalized to take values 0 and ¡À1. Here $\mathcal{L}_2$ corresponds to the Gauss-Bonnet term. Varying the action in Eq.~(\ref{action}) with respect to the metric tensor $g_{\mu\nu}$ we obtain the Lovelock field equations written in terms of generalized Einstein tensor \cite{MyersPRD1988}
\begin{equation}
G_{\mu\nu}\equiv -\frac{c_0}{2}g_{\mu\nu}+G_{\mu\nu}^{E}+\mathcal{G}_{\mu\nu}=T_{\mu\nu}, \label{Einstein eq}
\end{equation}
where
\begin{equation}
G_{\mu\nu}=\frac{2}{\sqrt{-g}}\frac{\delta S_L}{\delta g^{\mu\nu}},~~T_{\mu\nu}=-\frac{2}{\sqrt{-g}}\frac{\delta S_M}{\delta g^{\mu\nu}}.
\end{equation}
$T_{\mu\nu}$ is the energy momentum tensor associated with the electrostatic field and the dark fluid. $G_{\mu\nu}^{E}$ is the Einstein tensor, i.e.
\begin{equation}
G_{\mu\nu}^{E}=R_{\mu\nu}-\frac{1}{2}g_{\mu\nu}R.
\end{equation}
The Riemann-Lovelock curvature tensor $\mathcal{G}_{\mu\nu}$ is given as \cite{MyersPRD1988}
\begin{equation}
\mathcal{G}_\mu{}^{\nu}=\sum\limits^{m}_{k=2}c_k\delta _{\mu c_{1}\cdots c_{k}\beta_{1}\cdots\beta _{k}}^{\nu\mu _{1}\cdots\mu_{k}\nu _{1}\cdots\nu_{k}}R^{\alpha _{1}\beta _{1}}{}_{\mu _{1}\nu _{1}}\cdots R^{\alpha _{k}\beta _{k}}{}_{\mu _{k}\nu _{k}}.
\end{equation}
It is observed that the field equations are of second order as the Lovelock gravity is the sum of dimensionally continued Euler densities.

\section{Electrostatic field and dark fluid with a nonlinear equation of state in D-dimensional spacetime}
\label{section3}

The Lagrangian of electromagnetic field reads $-\frac{1}{2}F_{\mu\nu}F^{\mu\nu}$, and the vacuum Maxwell equations are derived as
\begin{align}
\partial_{\mu}(\sqrt{-g}F_{\mu\nu})=0,
\label{Maxwell}
\end{align}
where $F_{\mu\nu}=\partial_{\mu}A_{\nu}-\partial_{\nu}A_{\mu}$ is the electromagnetic tensor. Using Eq.~(\ref{Maxwell}), the gauge potential is obtained as follows
\begin{align}
A_{\mu}=\frac{-Q}{(D-3)r^{D-3}}\delta^{0}_{\mu},
\label{gaugeA}
\end{align}
where $Q$ is an integration constant which is related to the electric charge. It is easy to show that the nonzero components of the electromagnetic tensor are
\begin{align}
F_{tr}=F_{rt}=\frac{Q}{r^{D-2}}.
\label{emtensor}
\end{align}
The energy-momentum tensor of electromagnetic field is
\begin{align}
T^E_{\mu\nu}=2\left(F_{\mu\rho}{F_{\nu}}^{\rho}-\frac{1}{4}g_{\mu\nu}F_{\rho\sigma}F^{\rho\sigma}\right),
\label{EMTemf}
\end{align}
and its nonzero components, considering the electrostatic field case, are
\begin{align}
{{T^E}_t}^{t}&={{T^E}_r}^{r}=-\frac{Q^2}{r^{2(D-2)}},\\
{{T^E}_{\theta_1}}^{\theta_1}&={{T^E}_{\theta_2}}^{\theta_2}=\cdots={{T^E}_{\theta_{D-2}}}^{\theta_{D-2}}=\frac{Q^2}{r^{2(D-2)}},
\end{align}
where $\theta_i$($i=1,2,\cdots,D-2$) are the $D-2$ variables which describe the space-time section with constant curvature.

We then study the dark fluid with a nonlinear equation of state(EoS) $p_d=-\frac{B}{\rho_d}$, where $B$ is a positive constant. For $D$-dimensional spherically-symmetric spacetime, the energy momentum tensor of the dark fluid can be written as
\begin{equation}
{T^d}_t{}^t=\chi(r),~~{T^d}_t{}^i=0,~~{T^d}_i{}^j=\xi(r)\frac{r_ir^j}{r_nr^n}+\eta(r)\delta_i{}^j.\label{EMTdf1}
\end{equation}
Since we are considering static spherically-symmetric spacetime, the $r-r$ component of the energy-momentum tensor should be equal to the $t-t$ component, i.e.,
\begin{align}
{{T^d}_t}^t={{T^d}_r}^r=-\rho_d(r),
\label{EMTdftr}
\end{align}
If one takes isotropic average over the angles,
\begin{equation}
\langle r_i r^j\rangle=\frac{r_n r^n\delta_i{}^j}{(D-1)},\label{rAverage}
\end{equation}
one obtains
\begin{equation}
\langle {T^d}_i{}^j\rangle=\left(\frac{\xi(r)}{D-1}+\eta(r)\right)\delta_i{}^j=p_{d}(r)\delta_i{}^j=-\frac{B}{\rho_d(r)}\delta_i{}^j.\label{Tdfaverage}
\end{equation}
Considering Eqs.~(\ref{EMTdftr}) and (\ref{Tdfaverage}), $\xi(r)$ and $\eta(r)$ should be expressed as
\begin{equation}
\xi(r)=\alpha_1\rho_d(r)+\frac{\beta_1}{\rho_d(r)},~~\eta(r)=\alpha_2\rho_d(r)+\frac{\beta_2}{\rho_d(r)},\label{xieta}
\end{equation}
with the parameters $\alpha_i$ and $\beta_i$, constrained by Eqs.~(\ref{EMTdftr}) and (\ref{Tdfaverage}), yielding
\begin{align}
\alpha_1=-\frac{D-1}{D-2},~~\alpha_2=\frac{1}{D-2},~~\beta_1=\frac{D-1}{D-2}B,~~\beta_2=-\frac{D-1}{D-2}B.
\end{align}
Thus the angular components of the energy-momentum tensor are obtained as
\begin{align}
{{T^d}_{\theta_1}}^{\theta_1}&={{T^d}_{\theta_2}}^{\theta_2}=\cdots={{T^d}_{\theta_{D-2}}}^{\theta_{D-2}}=\frac{1}{D-2}\rho_d(r)-\frac{(D-1)B}{(D-2)\rho_d(r)}.
\end{align}
Along with the electrostatic field, the total energy-momentum tensor components are
\begin{align}
{T_t}^t&={T_r}^r=-\frac{Q^2}{r^{2(D-2)}}-\rho_d(r),\label{Ttt}\\
{T_{\theta_1}}^{\theta_1}&={T_{\theta_2}}^{\theta_2}=\cdots={T_{\theta_{D-2}}}^{\theta_{D-2}}=\frac{Q^2}{r^{2(D-2)}}+\frac{1}{D-2}\rho_d(r)-\frac{(D-1)B}{(D-2)\rho_d(r)}.\label{Ttheta}
\end{align}

\section{Charged Lovelock black holes surrounded by non-linear-EoS-dark fluid}
\label{section4}

The metric for general static spherically-symmetric spacetime in $D$-dimensions can be written as
\begin{equation}
ds^2=-f(r)dt^2+\frac{1}{f(r)} dr^2+r^2 d\Omega^2_{D-2},\label{dsf}
\end{equation}
where $d\Omega_{D-2}$ is a line element on a $D-2$ dimensional hypersurface with constant scalar curvature $\kappa=1,0$ and $-1$, respectively for spherical, flat and hyperbolic spaces.

We introduce $F(r)$ as
\begin{equation}
f(r)=\kappa-r^2F(r),\label{fF}
\end{equation}
then the components of the generalized Einstein tensor defined in Eq.~(\ref{Einstein eq}) yield
\begin{align}
{G_t}^t&={G_r}^r=-\frac{D-2}{2}\left[r\frac{\mathrm{d}}{\mathrm{d}r}+(D-1)\right]P\left[F(r)\right],\label{Gtt}\\
{G_{\theta_1}}^{\theta_1}&={G_{\theta_2}}^{\theta_2}=\cdots={G_{\theta_{D-2}}}^{\theta_{D-2}}=-\left[\frac{(D-1)(D-2)}{2}+(D-1)r\frac{\mathrm{d}}{\mathrm{d}r}+\frac{1}{2}r^2\frac{\mathrm{d}}{\mathrm{d}r^2}\right]P\left[F(r)\right],\label{Gtheta}
\end{align}
where $P\left[F(r)\right]$ is a polynomial of $F(r)$
\begin{equation}
P\left[F(r)\right]=\sum\limits^{m}_{k=0}\widehat{c}_k F^k(r), \label{PFr}
\end{equation}
with the coefficient $\widehat{c}_k$ defined by
\begin{eqnarray}\nonumber
\widehat{c}_0&=&\frac{c_0}{[(D-1)(D-2)]},~~~\widehat{c}_1=1,\\
\widehat{c}_k&=&\prod\limits_{i=3}^{2k}(D-i)c_k,~~(k>1).\nonumber
\end{eqnarray}

\begin{figure}[htb]
\centering
\includegraphics[width=4.0in]{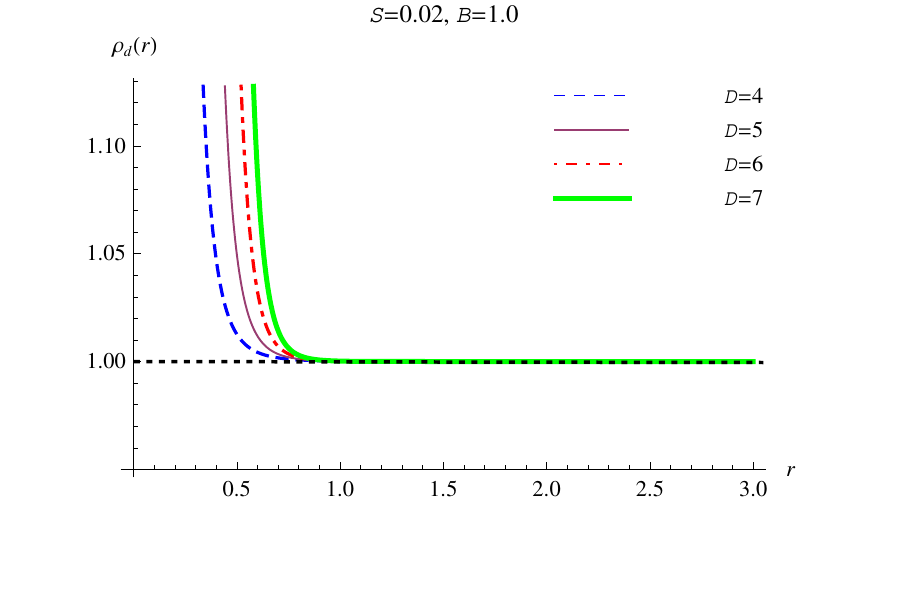}
\caption{Energy density of the dark fluid as function of the radial coordinate.}\label{darkfluid}
\end{figure}

Combining Eqs.~(\ref{Ttt}-\ref{Ttheta}) and~(\ref{Gtt}-\ref{Gtheta}), one finds the energy density of the dark fluid satisfying the following differential equation
\begin{equation}
r\rho_d'(r)+(D-1)\rho_d(r)-\frac{B(D-1)}{\rho_d(r)}=0, \label{eqdfdensity}
\end{equation}
then one obtains
\begin{equation}
\rho_d(r)=\sqrt{B+\frac{S^2}{r^{2(D-1)}}}, \label{dfdensity}
\end{equation}
where $S^2$ with $S>0$ is an integration constant. We observe that $\rho_d(r)\rightarrow\sqrt{B}$ when $r\rightarrow\infty$, which means that the dark fluid acts like a cosmological constant very far from the black hole, and it gathers more densely as it moves toward the black hole because of the gravitation, as displayed in Fig.~\ref{darkfluid}. It's interesting to note that Eq.~(\ref{dfdensity}) holds in any specific Lovelock gravity theory. If we replace a Lovelock theory with another one, $\rho_d(r)$ satisfies the same differential equation. In fact, the alternation of specific Lovelock theory doesn't affect the distribution of the surrounding spacetime field, whatever it is.

By solving ${G_t}^t={T_t}^t$, $F(r)$ satisfies the following polynomial equation
\begin{equation}
P\left[F(r)\right]=\frac{2M}{(D-2)\Sigma_{D-2}r^{D-1}}-\frac{2Q^2}{(D-2)(D-3)r^{2(D-2)}}+\frac{2\sqrt{B+\frac{S^2}{r^{2(D-1)}}}}{(D-1)(D-2)}-\frac{2S\mathrm{ArcSinh}\frac{S}{\sqrt{B}r^{D-1}}}{(D-1)(D-2)r^{D-1}}.\label{PFe}
\end{equation}
The thermodynamical quantities associated with black holes can be expressed in terms of a horizon radius $r_h$ which satisfies $f(r_h)=0$ in Eq.~(\ref{fF}), leading to
\begin{equation}
r_h^2=\frac{\kappa}{F(r_h)}.\label{rhF}
\end{equation}

The mass of the black hole in terms of the black hole horizon, by using the Eqs.~(\ref{PFe}) and (\ref{rhF}), reads
\begin{equation}
M=\frac{\Sigma_{D-2}}{2}\left[\sum\limits^m_{k=0}\frac{(D-2)\widehat{c}_k\kappa^k}{r_h^{-(D-2k-1)}}+\frac{2Q^2}{(D-3)r_h^{D-3}}-\frac{2\sqrt{B+\frac{S^2}{r_h^{2(D-1)}}}}{(D-1)r_h^{-(D-1)}}+\frac{2S\mathrm{ArcSinh}\frac{S}{\sqrt{B}r_h^{D-1}}}{D-1}\right].\label{M}
\end{equation}

Next we calculate thermodynamical quantities associated with black holes. The Hawking temperature associated with black holes is defined as $T=K/2\pi$, where $K$ is the surface gravity, which leads to
\begin{equation}
T=\frac{f^\prime(r)}{4\pi},
\end{equation}
which on Eq.~(\ref{fF}), reads
\begin{equation}
T=\frac{1}{4\pi \mathcal{N}(r_h)}\left[{\sum\limits^m_{k=0}\frac{(D-2k-1)\widehat{c}_k\kappa^{k}}{r_h^{2k}}}-\frac{2Q^2}{(D-2)r_h^{2(D-2)}}-\frac{2}{D-2}\sqrt{B+\frac{S^2}{r_h^{2(D-1)}}}\right],\label{T}
\end{equation}
with $$\mathcal{N}(r_h)={\sum\limits^m_{k=1}\frac{\widehat{c}_k\kappa^{k-1}k}{r_h^{2k-1}}}.$$

The entropy is given by
\begin{equation}
S=\int T^{-1}dM=\int T^{-1}\frac{dM}{dr_h}dr_h.
\end{equation}
Inserting Eq.~(\ref{T}) and the derivative of Eq.~(\ref{M}) into the above equation, the entropy becomes
\begin{equation}
S=2\pi(D-2)\Sigma_{D-2}\sum\limits^m_{k=1}\frac{\widehat{c}_k\kappa^{k-1}k}{(D-2k)r_h^{-(D-2k)}}.\label{S}
\end{equation}
which suggests that in general the Lovelock black hole surrounded by the dark fluid does not obey the area law.

The heat capacity is defined as
\begin{equation}
C=\frac{d M}{d T}=\frac{d M}{dr_h}\frac{dr_h}{dT}.\label{Ct}
\end{equation}
Thus using Eqs.~(\ref{M}) and (\ref{T}),
\begin{equation}
C=2\pi(D-2)\Sigma_{D-2}\frac{\mathcal{X}\mathcal{Y}^2}{\mathcal{Z}},\label{C}
\end{equation}
where
\begin{eqnarray}\nonumber
\mathcal{X}&=&\sum\limits^m_{k=0}(D-2k-1)\frac{\widehat{c}_k\kappa^{k}}{r^{-(D-2k-2)}_h}-\frac{2Q^2}{(D-2)r_h^{D-2}}-\frac{2\sqrt{B+\frac{S^2}{r_h^{2(D-1)}}}}{(D-2)r_h^{-(D-2)}},~~
\mathcal{Y}=\sum\limits^m_{k=1}\frac{\widehat{c}_k\kappa^{k-1}k}{r^{-(D-2k-1)}_h}, \\
\mathcal{Z}&=&\sum\limits^m_{k,s}(D-2s-1)(2k-2s-1)\frac{\widehat{c}_k\widehat{c}_s\kappa^{k+s-1}k}{r^{-2(D-k-s-2)}_h}+\frac{2Q^2}{D-2}\sum\limits^m_{k=1}(2D-2k-3)\frac{\widehat{c}_k\kappa^{k-1}k}{r_h^{2k}} \nonumber\\
&&+\frac{2}{D-2}\sum\limits^m_{k=1}[B(1-2k)r^{2D}+S^2(D-2k)r^2]\frac{\widehat{c}_k\kappa^{k-1}k}{r^{2(k+2)}_h}\frac{1}{\sqrt{B+\frac{S^2}{r_h^{2(D-1)}}}}.
\end{eqnarray}
Thermodynamical stability of black holes is directly related to the sign of the heat capacity. The positivity of heat capacity indicates that a thermodynamic system is stable whereas its negativity implies that a thermodynamic system is unstable.

In the following, we discuss charged black holes surrounded by nonlinear-EoS-dark fluid in D-dimensional Einstein gravity and Gauss-Bonnet gravity, as two special cases.
\subsection{D-dimensional Einstein black hole}
\begin{figure}[htb]
\begin{minipage}[t]{0.33\linewidth}
\centering
\includegraphics[width=2.3in]{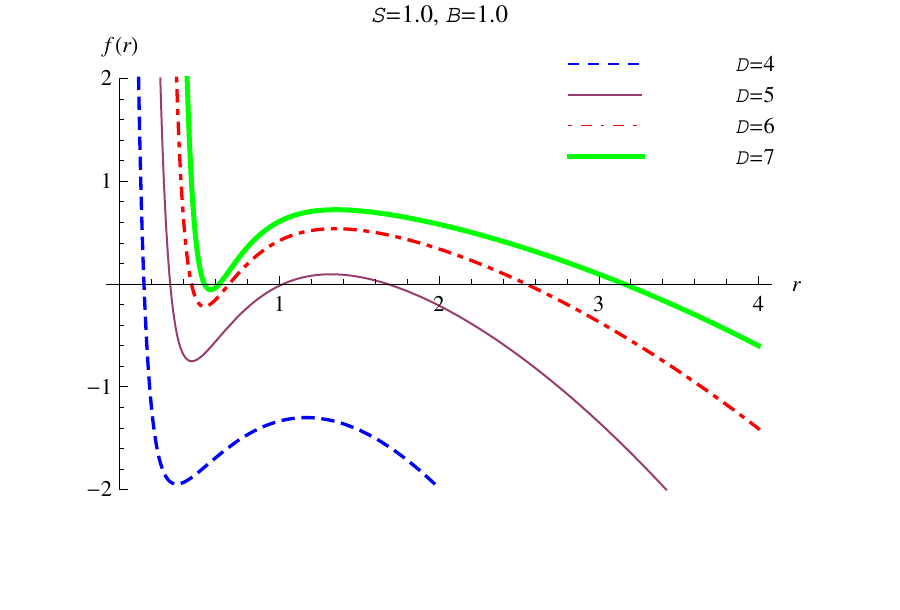}
\label{fig:side:a}
\end{minipage}%
\begin{minipage}[t]{0.33\linewidth}
\centering
\includegraphics[width=2.3in]{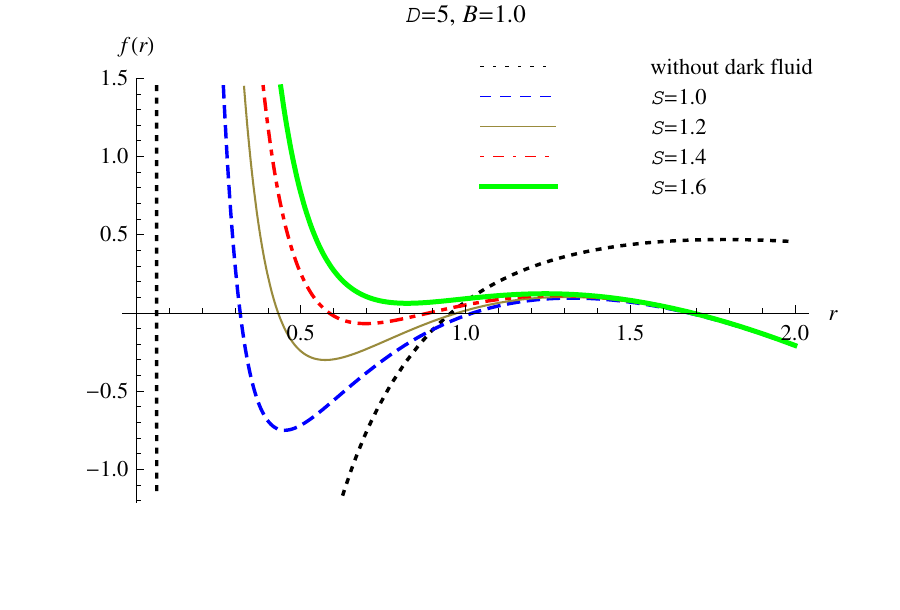}
\label{fig:side:b}
\end{minipage}
\begin{minipage}[t]{0.33\linewidth}
\centering
\includegraphics[width=2.3in]{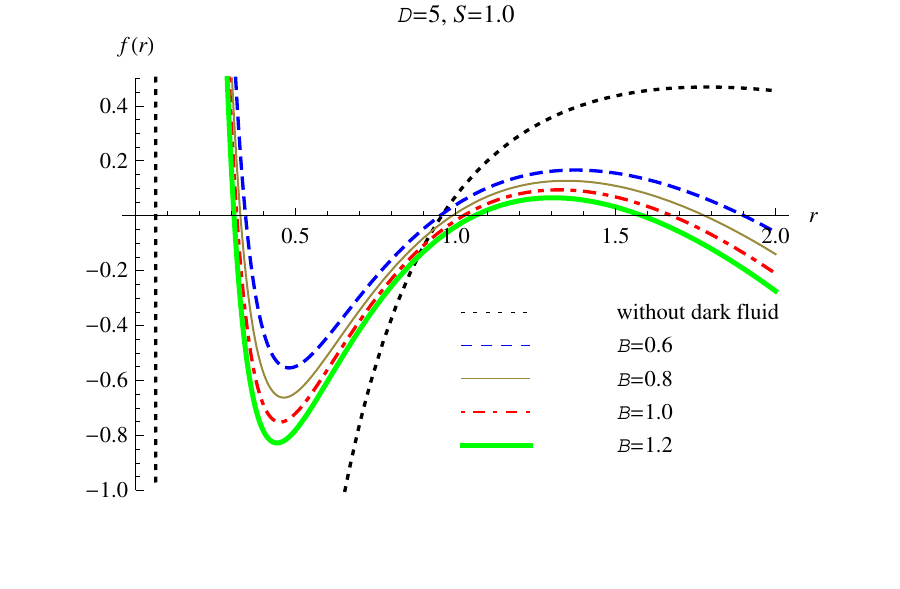}
\label{fig:side:b}
\end{minipage}
\caption{The function $f(r)$ in $D$-dimensional Einstein gravity for different values of $D$, $S$ and $B$, where we set $\widetilde{M}=1$, $Q=0.1$ and $c_0=-1$.}\label{frvaryDSB}
\end{figure}

In general, for $D\geq4$ and $\kappa=1$, $c_k=0$ for $k\geq2$, we obtain the metric function
\begin{eqnarray}
f(r)&=&1-\frac{16\pi\widetilde{M}}{(D-2)\sum_{D-2}r^{D-3}}+\frac{2Q^2}{(D-2)(D-3)r^{2(D-3)}}-\frac{2\sqrt{B+\frac{S^2}{r^{2(D-1)}}}r^{2}}{(D-1)(D-2)} \nonumber \\
&&+\frac{2S\mathrm{ArcSinh}\frac{S}{\sqrt{B}r^{D-1}}}{(D-1)(D-2)r^{D-3}}+\frac{c_0r^2}{(D-1)(D-2)},\label{Einsteinfr}
\end{eqnarray}
where $\widetilde{M}=\frac{M}{8\pi}$ with $M$ considered as the mass of a black hole, and it reduces to
\begin{equation}
f(r)=1-\frac{2\widetilde{M}}{r}+\frac{Q^2}{r^{2}}-\frac{r^2}{3}\sqrt{B+\frac{S^2}{r^6}}+\frac{S}{3r}\mathrm{ArcSinh}\frac{S}{\sqrt{B}r^3}+\frac{c_0r^2}{6} \label{Einsteinfr2},
\end{equation}
when $D=4$. We observe that $f(r)\rightarrow1+\frac{(c_0-2\sqrt{B})r^2}{(D-1)(D-2)}$ when $r\rightarrow+\infty$, indicating that the asymptotic behavior of $f(r)$ at infinity is determined by $c_0-2\sqrt{B}$. For general dimensions, we plot the $f(r)$ function with varying values of $D$, $S$ and $B$ in Fig.~\ref{frvaryDSB}. One can observe that, the parameter $S$ is liable to affect the existence and position of the black hole horizon, while the parameter $B$ is liable to govern the existence and position of the cosmological horizon. The solutions can be classified into two types by horizon existence. The first is a black hole solution which has an inner horizon and a black hole horizon. The second solution is a globally naked solution, which does not have a black hole horizon but has a locally naked singularity.

\begin{figure}[htb]
\begin{minipage}[t]{0.33\linewidth}
\centering
\includegraphics[width=2.3in]{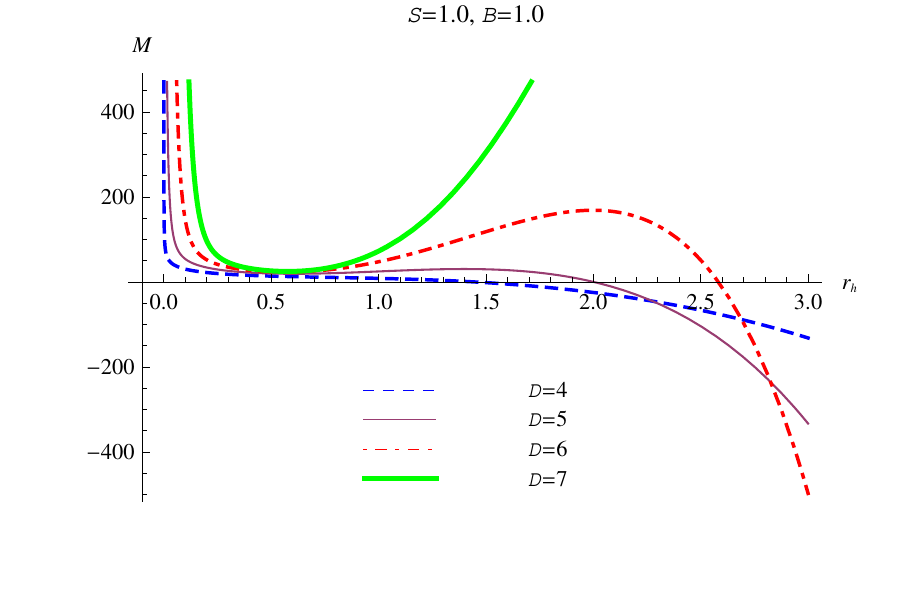}
\label{fig:side:a}
\end{minipage}%
\begin{minipage}[t]{0.33\linewidth}
\centering
\includegraphics[width=2.3in]{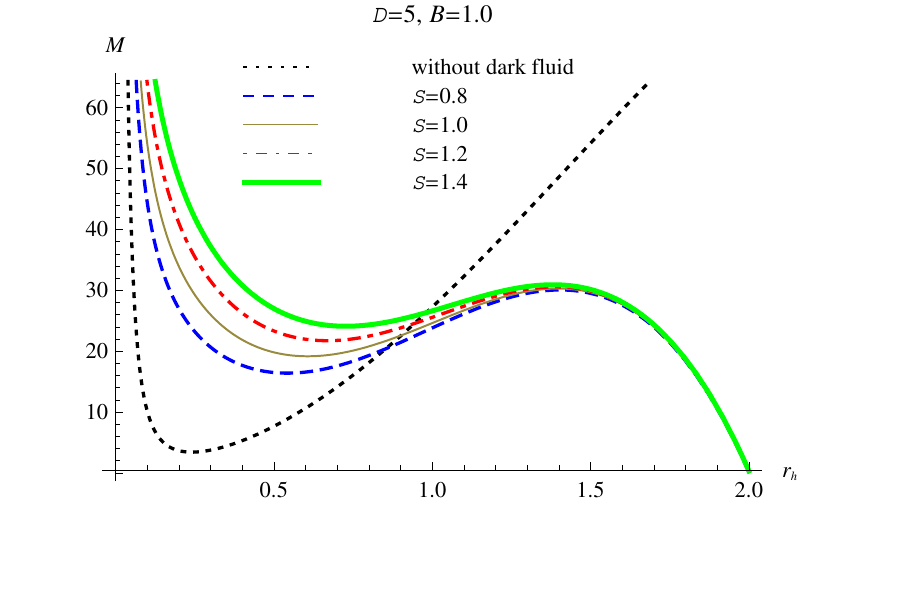}
\label{fig:side:b}
\end{minipage}
\begin{minipage}[t]{0.33\linewidth}
\centering
\includegraphics[width=2.3in]{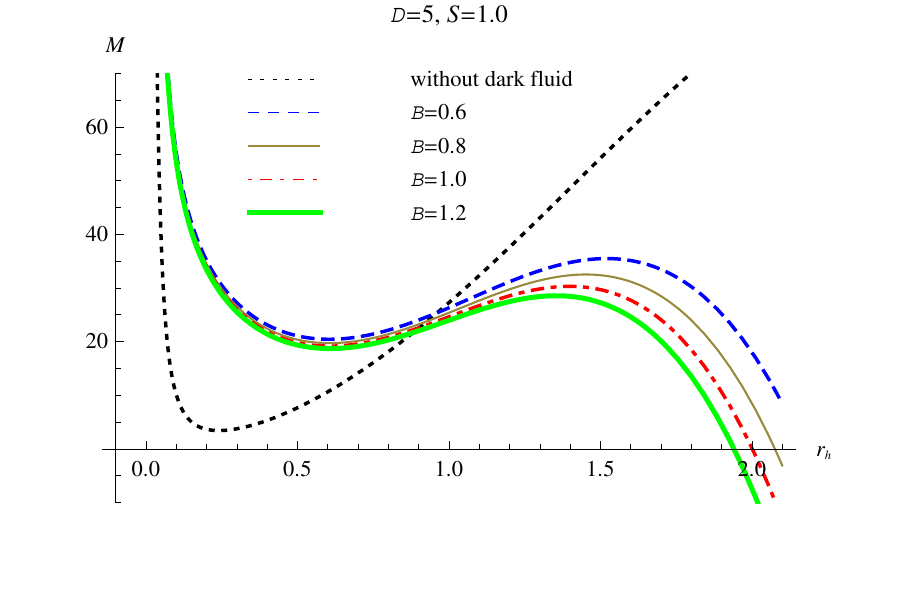}
\label{fig:side:b}
\end{minipage}
\caption{The mass function $M$ of $D$-dimensional Einstein black hole for different values of $D$, $S$ and $B$, where we set $Q=0.1$ and $c_0=-1$.}\label{MvaryDSB}
\end{figure}

For $D\geq4$, the mass of the black hole can be written as
\begin{equation}
M=\frac{\Sigma_{D-2}}{2}\left[\frac{c_0}{D-1}r_h^{D-1}+(D-2)r_h^{D-3}+\frac{2Q^2}{(D-3)r_h^{D-3}}-\frac{2\sqrt{B+\frac{S^2}{r_h^{2(D-1)}}}}{(D-1)r_h^{-(D-1)}}+\frac{2S\mathrm{ArcSinh}\frac{S}{\sqrt{B}r_h^{D-1}}}{D-1}\right],\label{MEinstein}
\end{equation}
which is plotted in Fig.~\ref{MvaryDSB}.

\begin{figure}[htb]
\begin{minipage}[t]{0.33\linewidth}
\centering
\includegraphics[width=2.3in]{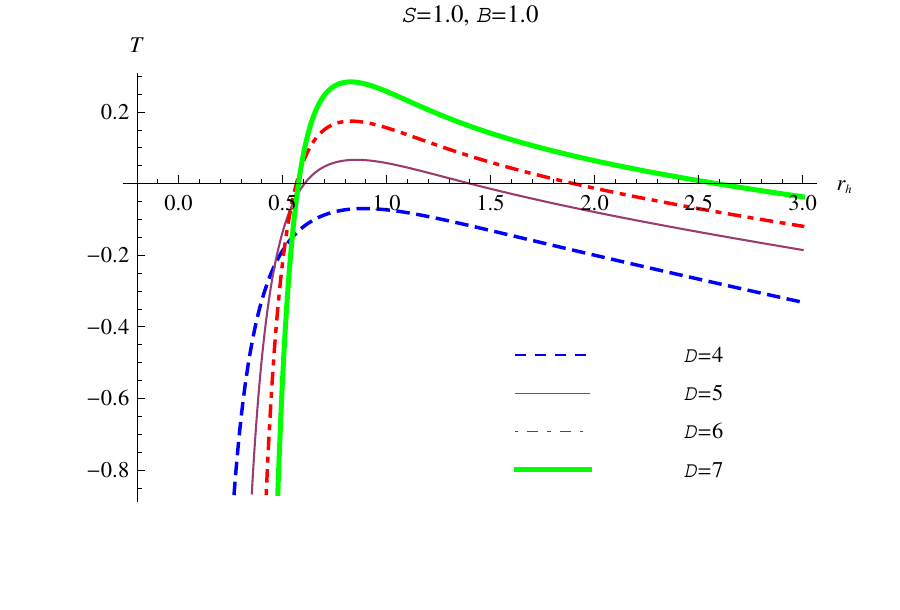}
\label{fig:side:a}
\end{minipage}%
\begin{minipage}[t]{0.33\linewidth}
\centering
\includegraphics[width=2.3in]{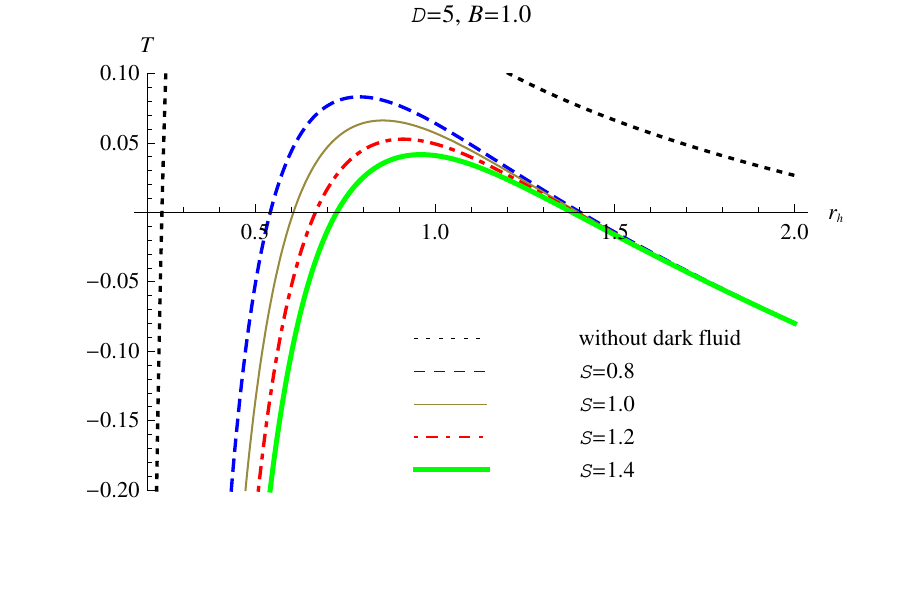}
\label{fig:side:b}
\end{minipage}
\begin{minipage}[t]{0.33\linewidth}
\centering
\includegraphics[width=2.3in]{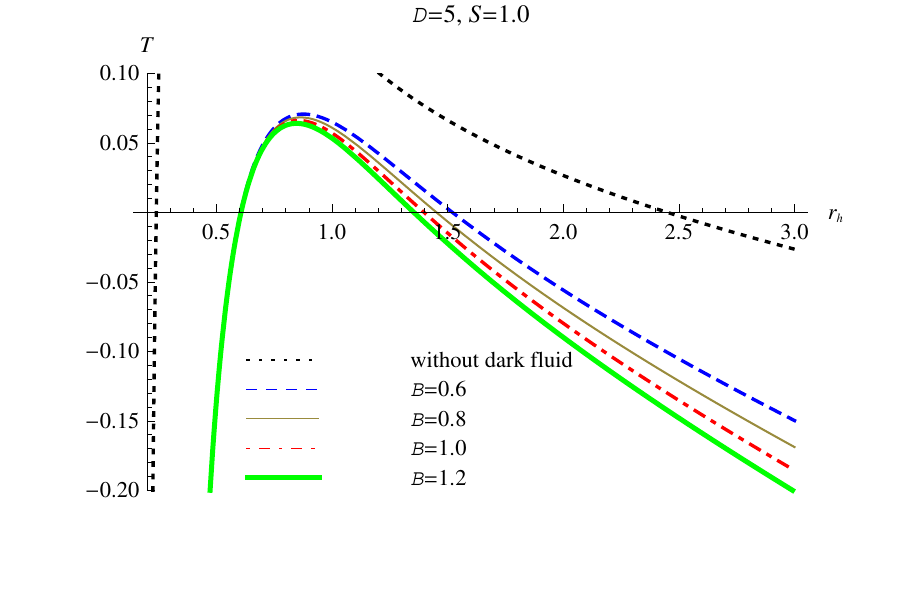}
\label{fig:side:b}
\end{minipage}
\caption{The Hawking temperature $T$ of $D$-dimensional Einstein black hole for different values of $D$, $S$ and $B$, where we set $Q=0.1$ and $c_0=-1$.}\label{TvaryDSB}
\end{figure}

The Hawking temperature is given by
\begin{equation}
T=\frac{r_h}{4\pi}\left[\frac{c_0}{D-2}+\frac{D-3}{r_h^2}-\frac{2Q^2}{(D-2)r_h^{2(D-2)}}-\frac{2}{D-2}\sqrt{B+\frac{S^2}{r_h^{2(D-1)}}}\right],\label{TEinstein}
\end{equation}
which is plotted in Fig.~\ref{TvaryDSB}.

\begin{figure}[htb]
\begin{minipage}[t]{0.33\linewidth}
\centering
\includegraphics[width=2.3in]{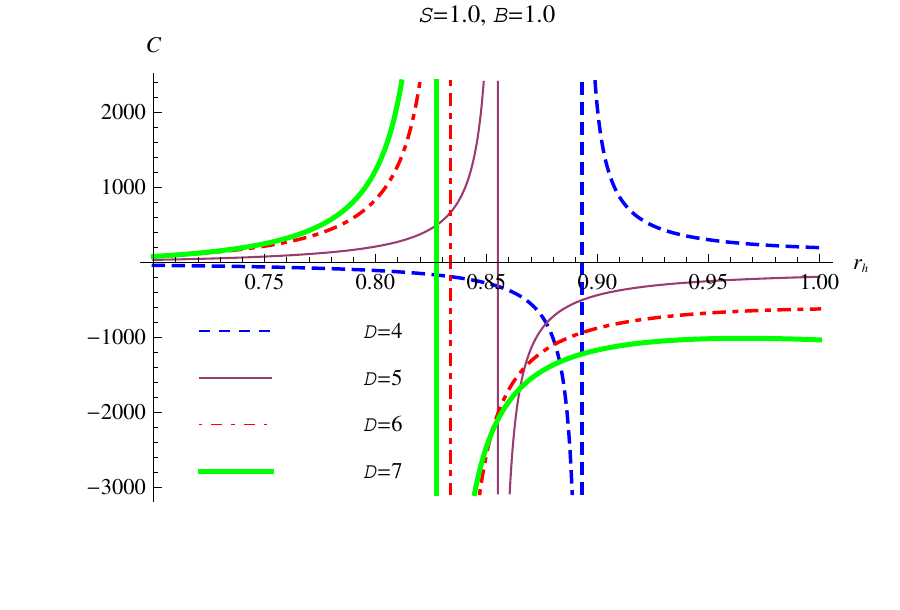}
\label{fig:side:a}
\end{minipage}%
\begin{minipage}[t]{0.33\linewidth}
\centering
\includegraphics[width=2.3in]{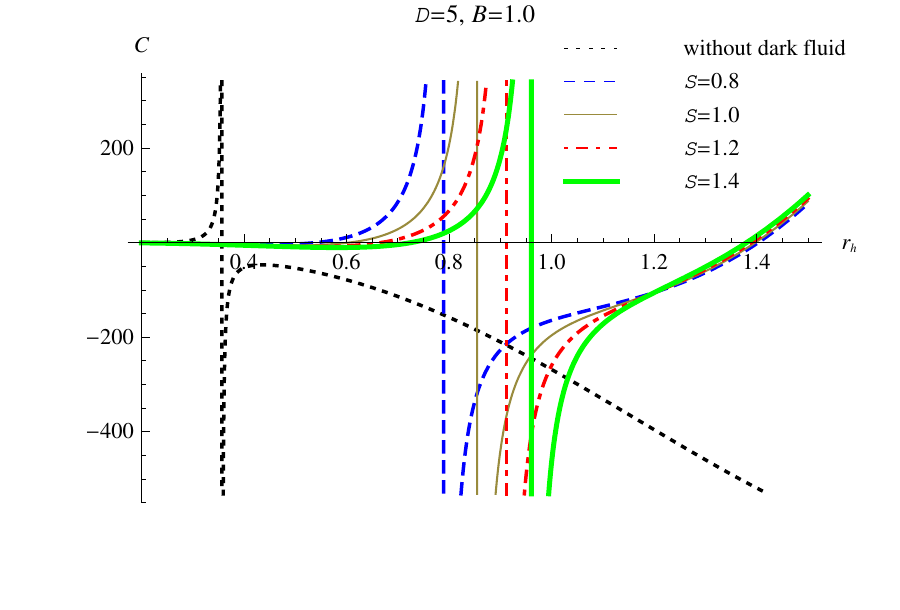}
\label{fig:side:b}
\end{minipage}
\begin{minipage}[t]{0.33\linewidth}
\centering
\includegraphics[width=2.3in]{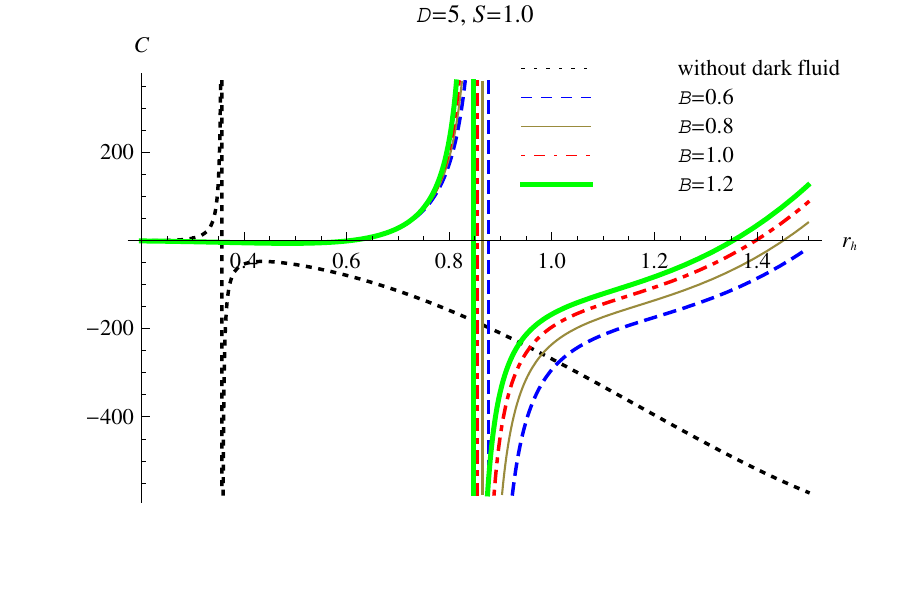}
\label{fig:side:b}
\end{minipage}
\caption{The heat capacity $C$ of $D$-dimensional Einstein black hole for different values of $D$, $S$ and $B$, where we set $Q=0.1$ and $c_0=-1$.}\label{CvaryDSB}
\end{figure}

In this case, the entropy is proportional to the area of the event horizon
\begin{equation}
S=2\pi\Sigma_{D-2}r_h^{D-2}.\label{S}
\end{equation}
For a black hole in D-dimensional spacetime, we can calculate the heat capacity using Eq.~(\ref{C}), with
\begin{eqnarray}
\mathcal{X}&=&\frac{c_0}{D-2}r_h^{D-2}+(D-3)r_h^{D-4}-\frac{2Q^2}{(D-2)r_h^{D-2}}-\frac{2\sqrt{B+\frac{S^2}{r_h^{2(D-1)}}}}{(D-2)r_h^{-(D-2)}},~~
\mathcal{Y}=r_h^{D-3}, \nonumber \\
\mathcal{Z}&=&\frac{c_0}{D-2}r_h^{2(D-3)}-(D-3)r_h^{2(D-4)}+\frac{2(2D-5)Q^2}{(D-2)r_h^2}-\frac{2}{(D-2)r_h^4}\big[Br_h^{2(D-1)} \nonumber\\
&&-(D-2)S^2\big]\frac{1}{\sqrt{B+\frac{S^2}{r_h^{2(D-1)}}}}.
\end{eqnarray}
We plot the $C$ function in Fig.~\ref{CvaryDSB}, and find that the heat capacity depends on the spacetime dimensions $D$, and also the dark fluid parameters $S$ and $B$. For given values of $D$, $S$ and $B$, the heat capacity diverges at some critical radius $r_c$, where the heat capacity changes its sign. As $S$ increases, $r_c$ gets right-shifted.

\subsection{Gauss-Bonnet black hole}
\begin{figure}[htb]
\begin{minipage}[t]{0.33\linewidth}
\centering
\includegraphics[width=2.3in]{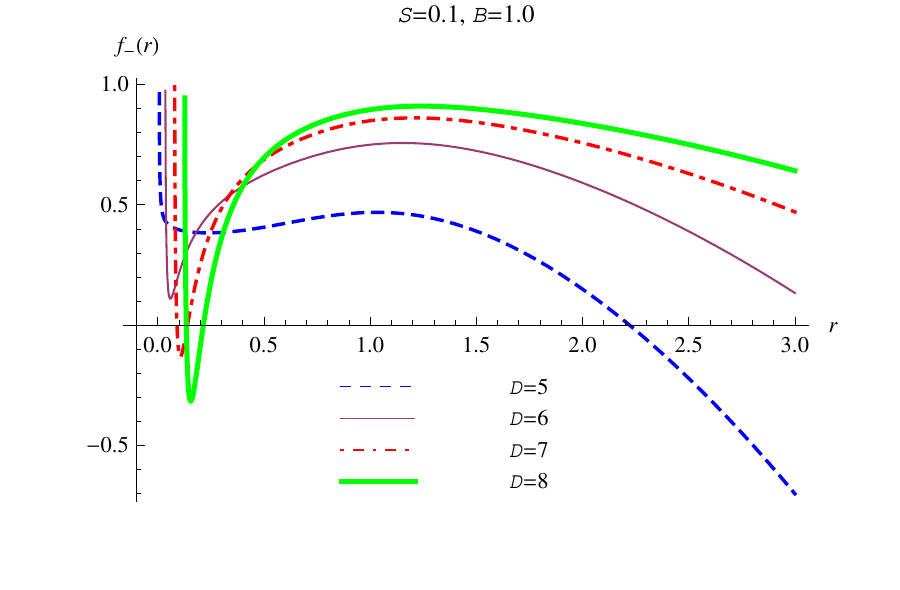}
\label{fig:side:a}
\end{minipage}%
\begin{minipage}[t]{0.33\linewidth}
\centering
\includegraphics[width=2.3in]{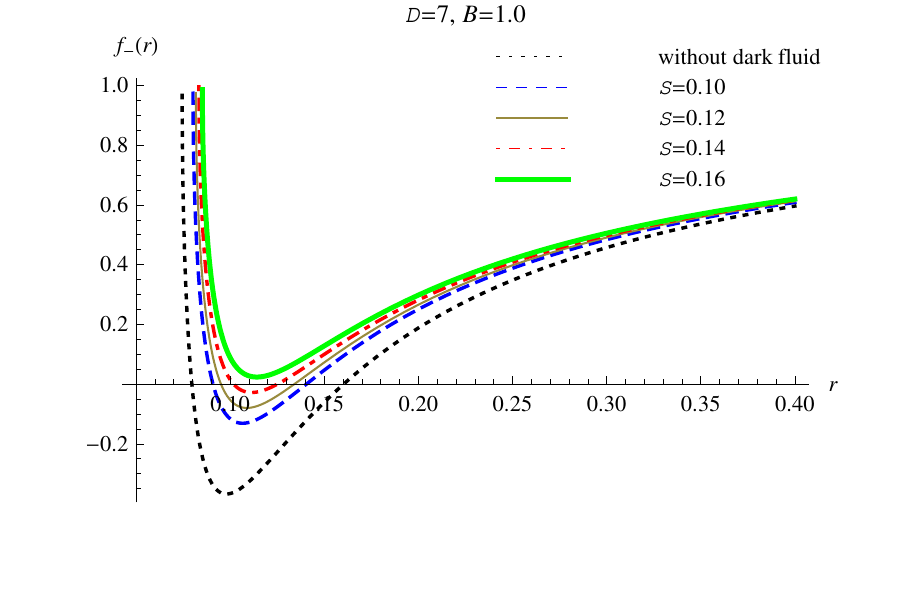}
\label{fig:side:b}
\end{minipage}
\begin{minipage}[t]{0.33\linewidth}
\centering
\includegraphics[width=2.3in]{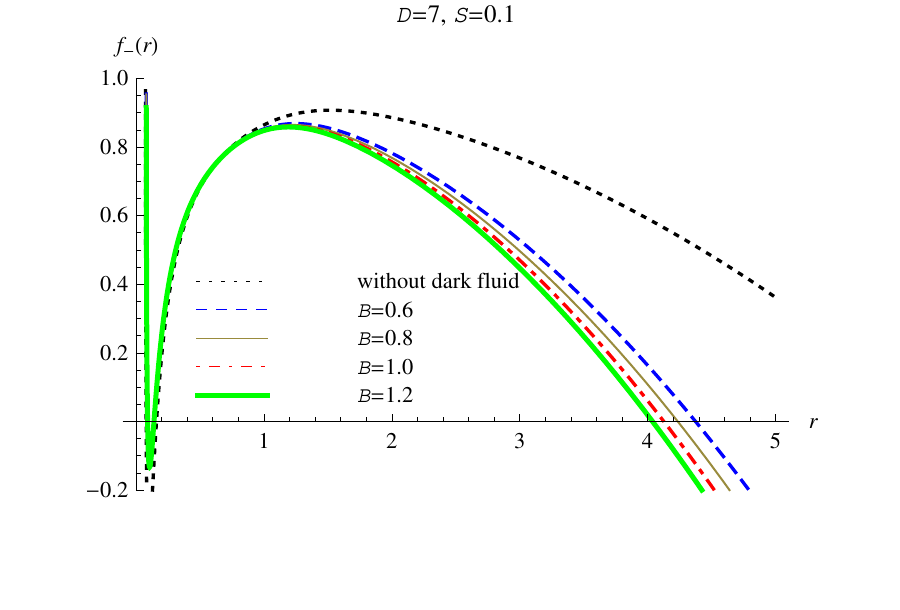}
\label{fig:side:b}
\end{minipage}
\caption{The $f_{-}(r)$ function in Gauss-Bonnet gravity for different values of $D$, $S$ and $B$, where we set $M=1$, $c_2=1$, $Q=0.01$ and $c_0=-1$.}\label{GBfrvaryDSB}
\end{figure}

The Gauss-Bonnet solution is obtained when $c_2\neq0$ and $c_k=0$ for $k\geq3$. Setting $\kappa=1$ and considering Eq.~(\ref{PFe}), we obtain two branches for the
spacetime solution, which are given by
\begin{equation}
f_\pm(r)=1+\frac{r^2}{2\widehat{c}_2}\times \left(1\pm\sqrt{\mathcal{F}(r)}\right)~\label{GBfr}
\end{equation}
with
\begin{eqnarray}
\mathcal{F}(r)&=&1-\frac{4c_0\widehat{c}_2}{(D-1)(D-2)}+\frac{64\pi\widehat{c}_2\widetilde{M}}{(D-2)\sum_{D-2}r^{D-1}}-\frac{8\widehat{c}_2Q^2}{(D-2)(D-3)r^{2(D-2)}}~\nonumber\\
&&+\frac{8\widehat{c}_2\sqrt{B+\frac{S^2}{r_h^{2(D-1)}}}}{(D-1)(D-2)}-\frac{8\widehat{c}_2S\mathrm{ArcSinh}\frac{S}{\sqrt{B}r^{D-1}}}{(D-1)(D-2)r^{D-1}},
\end{eqnarray}
where $\widehat{c}_2=(D-3)(D-4)c_2$. To study the asymptotic behavior of $f_\pm(r)$, we take $r\rightarrow\infty$ and find that
\begin{equation}
f_{\pm}(r) \rightarrow 1+\frac{r^2}{2\widehat{c}_2}\Bigg(1\pm\sqrt{1-\frac{4\widehat{c}_2(c_0-2\sqrt{B})}{(D-1)(D-2)}}\Bigg),
\end{equation}
which reveal that, given $\widehat{c}_2$ with positive value, $f_{+}(r)$ tends to anti-de Sitter spacetime, while $f_{-}(r)$ tends to anti-de Sitter spacetime when $c_0>2\sqrt{B}$ and de Sitter spacetime when $c_0<2\sqrt{B}$. Considering that a realistic physical metric should tend to the Schwarzschild metric when there is neither cosmological constant nor gravitational sources, except a central mass, the $f_{+}(r)$ solution has no physical interest. Since the metric solution must be real-valued, we should consider the domain values of $r$ such that the radicand in Eq.~(\ref{GBfr}), i.e. $\mathcal{F}(r)$ keeps non-negative. We note that, for given $M$ and $Q$, $\mathcal{F}(r)$ function reaches to zero value at $r=r_b$ which was referred as `branch singularity' in Refs.~\cite{Torii:2005xu,Torii:2005nh}, leading to $f_{-}(r)$ well defined at the interval [$r_h$,$+\infty$]. For $f_{-}(r)$ solutions with an inner horizon $r_i$ and a black hole horizon $r_h$, since $\mathcal{F}(r_{i,h})=(1+2\widehat{c}_2/r_{i,h}^2)^2$ whereas $\mathcal{F}(r_b)=0$, the branch singularity $r_b$ satisfies $r_b<r_i\leq r_h$. The $f_{-}(r)$ solution with varying values of $D$, $S$ and $B$ is plotted in Fig.~\ref{GBfrvaryDSB}. Again one can conclude that the existence and position of the black hole horizon is liable to be affected by the parameter $S$, while the existence and position of the cosmological horizon is liable to be controlled by the parameter $B$.

\begin{figure}[htb]
\begin{minipage}[t]{0.33\linewidth}
\centering
\includegraphics[width=2.3in]{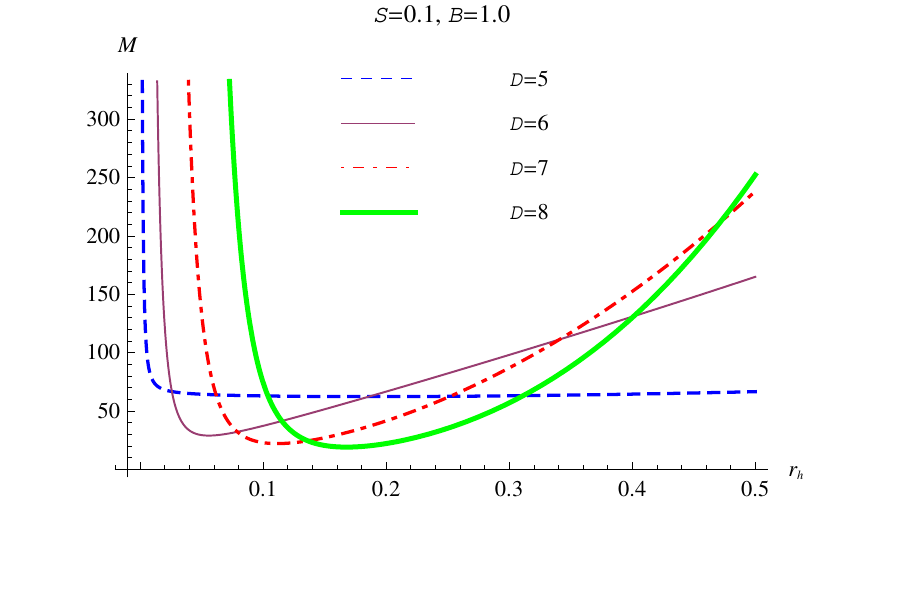}
\label{fig:side:a}
\end{minipage}%
\begin{minipage}[t]{0.33\linewidth}
\centering
\includegraphics[width=2.3in]{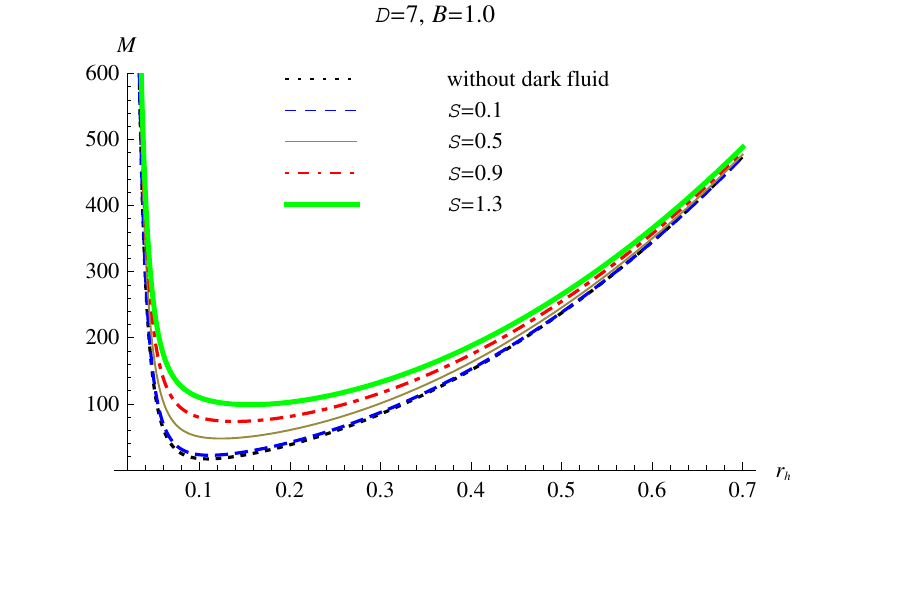}
\label{fig:side:b}
\end{minipage}
\begin{minipage}[t]{0.33\linewidth}
\centering
\includegraphics[width=2.3in]{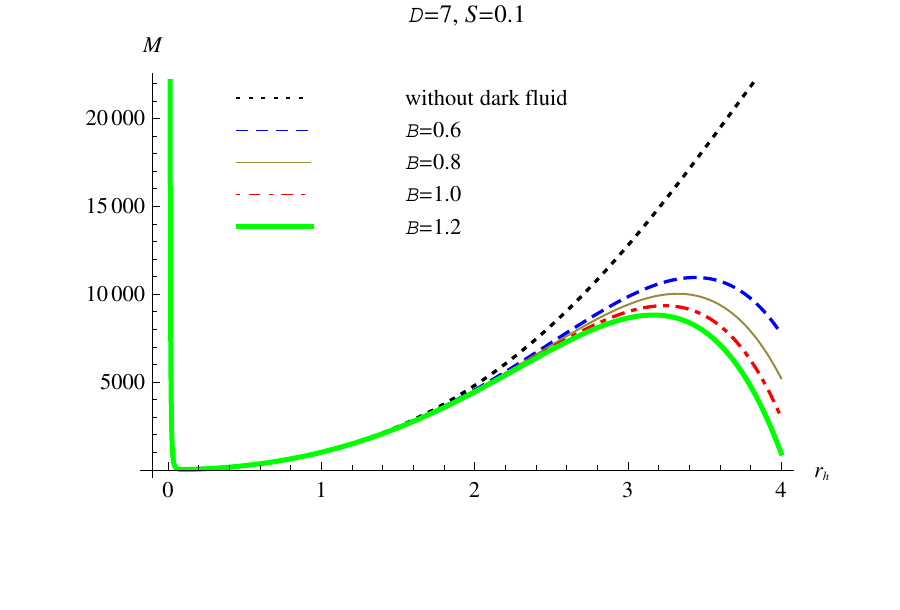}
\label{fig:side:b}
\end{minipage}
\caption{The mass function $M$ of Gauss-Bonnet black hole for different values of $D$, $S$ and $B$, where we set $c_2=1$, $Q=0.01$ and $c_0=-1$.}\label{GBMvaryDSB}
\end{figure}

The mass of the Gauss-Bonnet black hole is given by
\begin{eqnarray}
M&=&\frac{\sum_{D-2}}{2}\Bigg[(D-2)\widehat{c}_2r_h^{D-5}+\frac{c_0}{D-1}r_h^{D-1}+(D-2)r_h^{D-3}+\frac{2Q^2}{(D-3)r_h^{D-3}}-\frac{2\sqrt{B+\frac{S^2}{r_h^{2(D-1)}}}}{(D-1)r_h^{-(D-1)}}~\nonumber\\
&&+\frac{2S}{(D-1)}\mathrm{ArcSinh}\frac{S}{\sqrt{B}r_h^{D-1}}\Bigg],
\end{eqnarray}
which is plotted in Fig.~\ref{GBMvaryDSB}.

\begin{figure}[htb]
\begin{minipage}[t]{0.33\linewidth}
\centering
\includegraphics[width=2.3in]{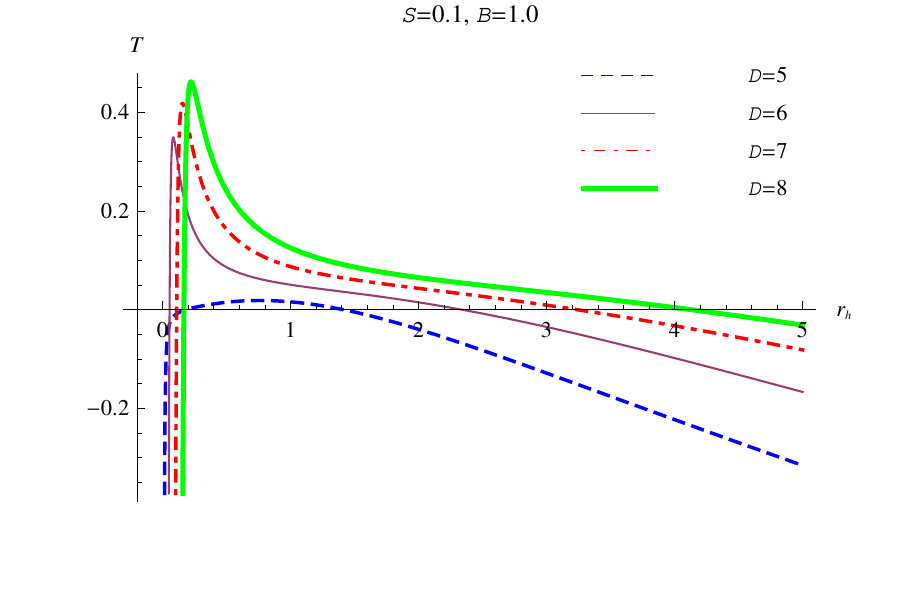}
\label{fig:side:a}
\end{minipage}%
\begin{minipage}[t]{0.33\linewidth}
\centering
\includegraphics[width=2.3in]{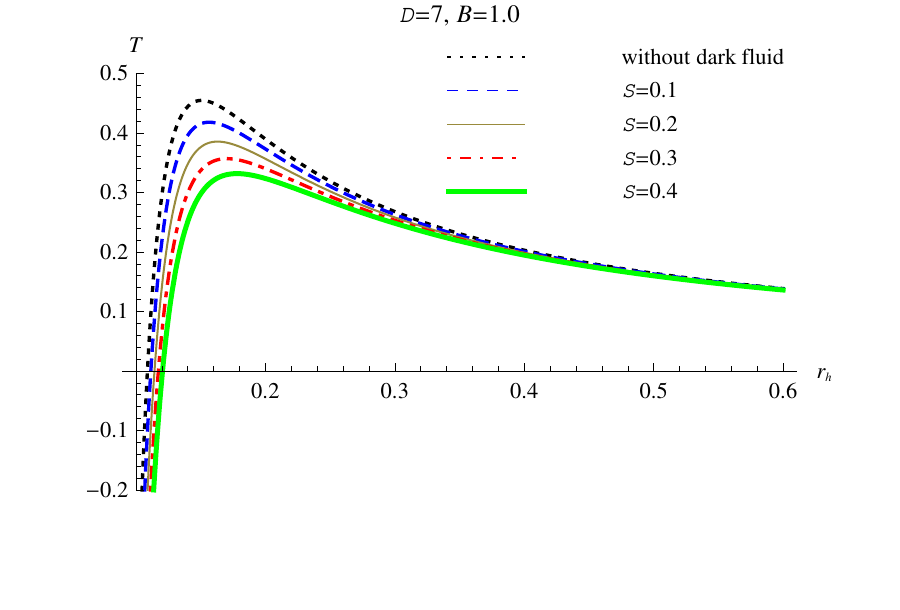}
\label{fig:side:b}
\end{minipage}
\begin{minipage}[t]{0.33\linewidth}
\centering
\includegraphics[width=2.3in]{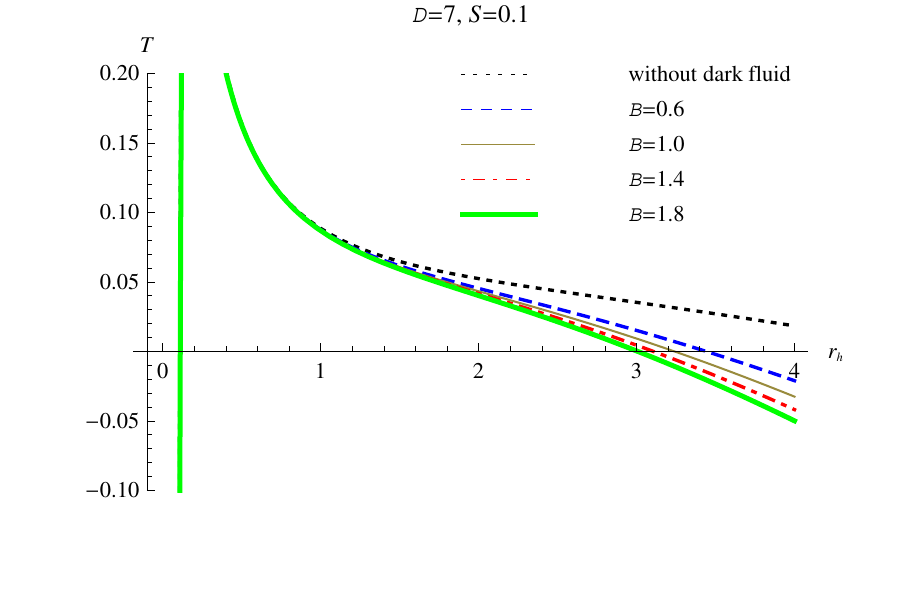}
\label{fig:side:b}
\end{minipage}
\caption{The Hawking temperature $T$ of Gauss-Bonnet black hole for different values of $D$, $S$ and $B$, where we set $c_2=1$, $Q=0.01$ and $c_0=-1$.}\label{GBTvaryDSB}
\end{figure}

The Hawking temperature is given by
\begin{equation}
T=\frac{r_h}{4\pi(1+2\widehat{c}_2r_h^{-2})}\left[\frac{c_0}{D-2}+\frac{D-3}{r_h^2}+\frac{(D-5)\widehat{c}_2}{r_h^4}-\frac{2Q^2}{(D-2)r_h^{D-2}}-\frac{2}{D-2}\sqrt{B+\frac{S^2}{r_h^{2(D-1)}}}\right],
\end{equation}
and its behaviors are displayed in Fig.~\ref{GBTvaryDSB}.

The entropy is
\begin{equation}
S=2(D-2)\pi\Sigma_{D-2}r_h^{D-4}\left[\frac{r_h^2}{D-2}+\frac{2\widehat{c}_2}{D-4}\right],
\end{equation}
showing that the area law is not fulfilled.

\begin{figure}[htb]
\begin{minipage}[t]{0.33\linewidth}
\centering
\includegraphics[width=2.3in]{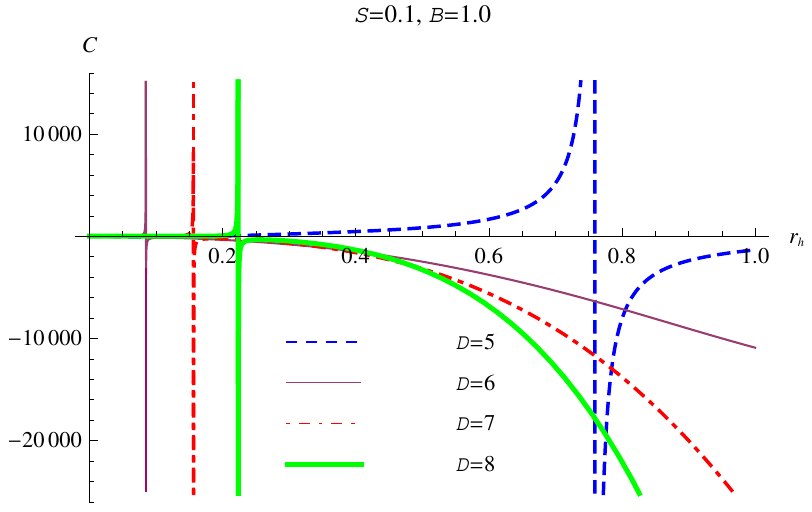}
\label{fig:side:a}
\end{minipage}%
\begin{minipage}[t]{0.33\linewidth}
\centering
\includegraphics[width=2.3in]{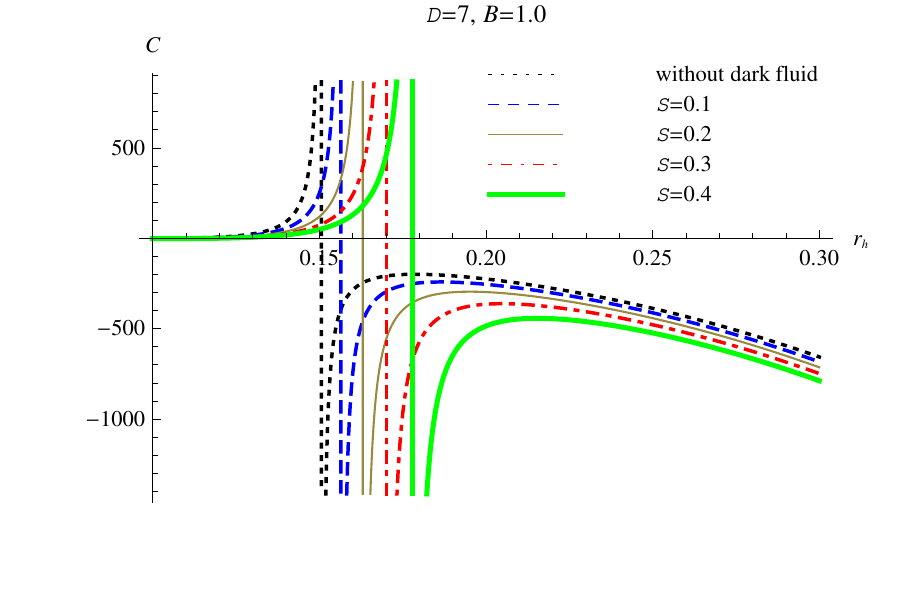}
\label{fig:side:b}
\end{minipage}
\begin{minipage}[t]{0.33\linewidth}
\centering
\includegraphics[width=2.3in]{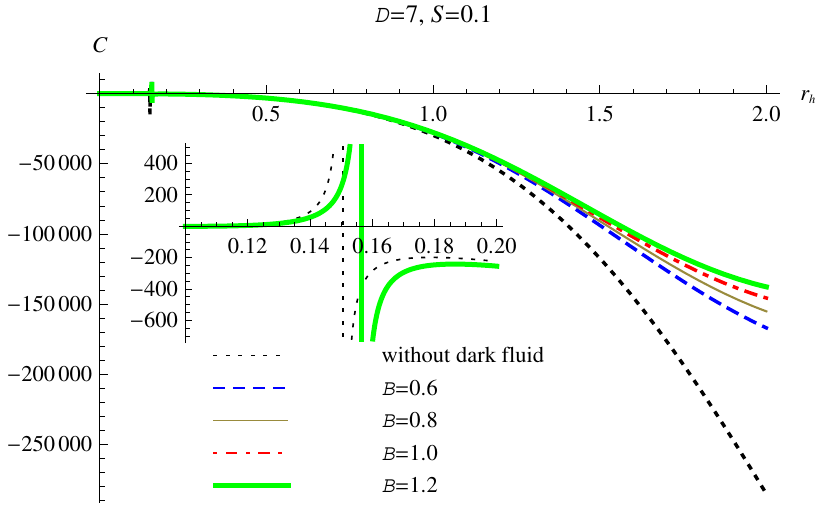}
\label{fig:side:b}
\end{minipage}
\caption{The heat capacity $C$ of Gauss-Bonnet black hole for different values of $D$, $S$ and $B$, where we set $c_2=1$, $Q=0.01$ and $c_0=-1$.}\label{GBCvaryDSB}
\end{figure}

In the case of Gauss-Bonnet gravity, the heat capacity is given by Eq.~(\ref{C}), where the parameters $\mathcal{X}$, $\mathcal{Y}$ and $\mathcal{Z}$ are
\begin{eqnarray}
\mathcal{X}&=&\frac{c_0}{D-2}r_h^{D-2}+(D-3)r_h^{D-4}+(D-5)\widehat{c}_2r_h^{D-6}-\frac{2Q^2}{(D-2)r_h^{D-2}}-\frac{2\sqrt{B+\frac{S^2}{r_h^{2(D-1)}}}}{(D-2)r_h^{-(D-2)}},~~\nonumber
\\
\mathcal{Y}&=&r_h^{D-3}+2\widehat{c}_2r_h^{D-5},~~\nonumber \\
\mathcal{Z}&=&\frac{c_0}{D-2}r_h^{2(D-3)}-(D-3)r_h^{2(D-4)}+\frac{6c_0\widehat{c}_2}{D-2}r_h^{2(D-4)}-(D-9)\widehat{c}_2r_h^{2(D-5)}~~\nonumber \\
&&-2(D-5){\widehat{c}_2}^2r_h^{2(D-6)}+\frac{2(2D-5)Q^2}{(D-2)r_h^2}+\frac{4(2D-7)\widehat{c}_2Q^2}{(D-2)r_h^4}-\frac{2}{(D-2)r_h^4}\big[Br^{2(D-1)} \nonumber\\
&&-(D-2)S^2\big]\frac{1}{\sqrt{B+\frac{S^2}{r_h^{2(D-1)}}}}-\frac{4\widehat{c}_2}{(D-2)r_h^6}\big[3Br^{2(D-1)}-(D-4)S^2\big]\frac{1}{\sqrt{B+\frac{S^2}{r_h^{2(D-1)}}}}.
\end{eqnarray}
The behavior of $C$ is represented by Fig.~\ref{GBCvaryDSB}. We conclude that there are regions which
are thermodynamically stable or unstable depending on
the dimensions of the spacetime and the dark fluid parameters. In the Gauss-Bonnet gravity case, there still exist critical horizon radius $r_c$ at which the heat capacity diverges, also $r_c$ is right-shifted by larger $S$.

\section{Concluding remarks}
\label{section5}

In this paper, we considered the static spherically-symmetric black hole in the presence of electrostatic field and Chaplygin-like dark fluid, in the framework of Lovelock gravity. By using the equation of state of the dark fluid $p_d=-B/\rho_d$, we obtained its energy density as function of the radial coordinate. We then solved the Lovelock gravitational equation and got a polynomial equation for $F(r)=(\kappa-f(r))/{r^2}$. According to this polynomial equation, the thermodynamical quantities of the black hole, as functions of the horizon radius, were calculated. Specially, we explicitly gave the metric solutions in the $D$-dimensional Einstein gravity case and the Gauss-Bonnet gravity case. We found that, for given $M$ and $Q$, $S$ was more likely to affect the existence and position of the black hole horizon while $B$ was liable to affect those of the cosmological horizon. In both cases, the black hole thermodynamical properties, with special
emphasis on the mass, Hawking temperature and heat capacity, were displayed by plots with varying $S$ and $B$. Concerning the stability of those black holes, we showed in figures the radial areas when heat capacity was negative and when it was positive, implying that the black holes were thermodynamically instable or stable.

\begin{figure}[htb]
\begin{minipage}[t]{0.5\linewidth}
\centering
\includegraphics[width=3.2in]{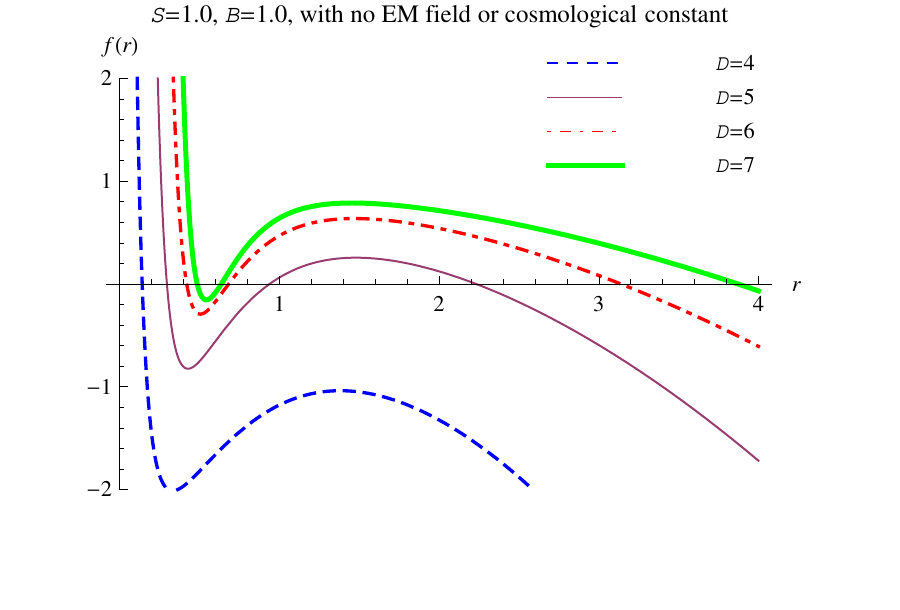}
\label{fig:side:b}
\end{minipage}
\begin{minipage}[t]{0.5\linewidth}
\centering
\includegraphics[width=3.2in]{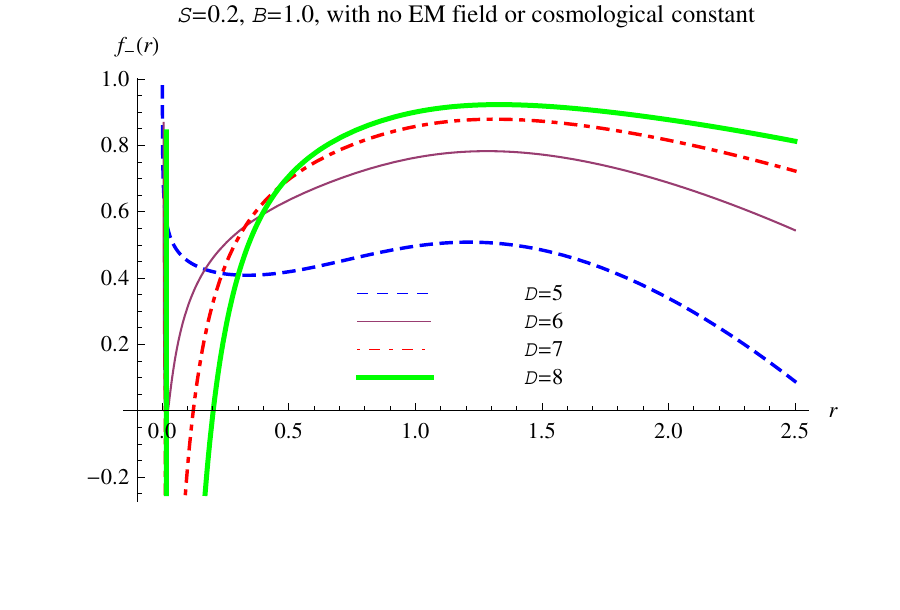}
\label{fig:side:b}
\end{minipage}
\caption{The $f(r)$ function in $D$-dimensional Einstein gravity and the $f_{-}(r)$ function in Gauss-Bonnet gravity when there is only dark fluid existed, i.e. $Q=0$ and $c_0=0$. We set $M=1, Q=0.1$ for the $f(r)$ function, and $M=1, Q=0.01$ for the $f_{-}(r)$ function.}\label{onlydarkfluid}
\end{figure}

Our study in this paper supposed the existences of the electrostatic field and cosmological constant. However it's interesting to consider the case with only dark fluid existed. To study the behavior of the dark fluid in this case, we first set $Q$ and $c_0$ in Eqs.~(\ref{Einsteinfr}) and~(\ref{GBfr}) to zero values, and then display $f(r)$ and $f_{-}(r)$ functions in Fig.~\ref{onlydarkfluid}. We find that the spacetime structure of the black holes in this case is similar to that of the Reissner-Nordstr$\ddot{o}$m-de Sitter black hole.

%\section*{Acknowledgments}


\begin{thebibliography}{10}
   \bibitem{reviewofGRtests} E. Asmodelle, {\it {Tests of General Relativity: A Review}}, [\href{http://arxiv.org/abs/1705.04397}{{\tt arXiv:gr-qc/1705.04397}}].
   \bibitem{stringtheory} B. A. Campbell, M. J. Duncan, N. Kaloper and K. A. Olive, \emph{Gravitational dynamics with lorentz chern-simons terms}, Nucl. Phys. B \textbf{351}, 778 (1991); I. Antoniadis, J. Rizos, and K. Tamvakis, \emph{Singularity-free cosmological solutions of the superstring effective action}, Nucl. Phys. B \textbf{415}, 497 (1994).
   \bibitem{branecosmology} L.~Randall and R.~Sundrum, \emph{A Large mass hierarchy from a small extra dimension}, Phys.\ Rev.\ Lett.\  {\bf 83}, 3370 (1999), [\href{http://arxiv.org/abs/hep-ph/9905221}{{\tt arXiv:hep-ph/9905221}}]; G.~R.~Dvali, G.~Gabadadze and M.~Porrati, \emph{4-D gravity on a brane in 5-D Minkowski space}, Phys.\ Lett.\ B {\bf 485}, 208 (2000), [\href{http://arxiv.org/abs/hep-th/0005016}{{\tt arXiv:hep-th/0005016}}]; G.~R.~Dvali and G.~Gabadadze, \emph{Gravity on a brane in infinite volume extra space}, Phys.\ Rev.\ D {\bf 63}, 065007 (2001), [\href{http://arxiv.org/abs/hep-th/0008054}{{\tt arXiv:hep-th/0008054}}].
   \bibitem{Lovelock} D. Lovelock, \emph{The Einstein tensor and its generalizations}, J. Math. Phys. \textbf{12}, 498 (1971); \emph{The Four-Dimensionality of Space and the Einstein Tensor}, J. Math. Phys. \textbf{13}, 874 (1972).
   \bibitem{freeghost} B. Zwiebach, \emph{Curvature squared terms and string theories}, Phys. Lett. B \textbf{156}, 315 (1985).
   \bibitem{GBterm}  D. J. Gross and J. H. Sloan, \emph{The quartic effective action for the heterotic string}, Nucl. Phys. B \textbf{291}, 41 (1987).
   \bibitem{BoulwarePRL1985} D. G. Boulware, S. Deser, \emph{String-Generated Gravity Models}, Phys. Rev. Lett. \textbf{55}, 2656 (1985).
   \bibitem{WheelerNPB1986} J.~T.~Wheeler, \emph{Symmetric Solutions to the Gauss-Bonnet Extended Einstein Equations}, Nucl.\ Phys.\ B {\bf 268}, 737 (1986).
   \bibitem{MyersPRD1988} R. C. Myers, J. Z. Simon, \emph{Black-hole thermodynamics in Lovelock gravity}, Phys. Rev. D {\bf{38}}, 2434 (1988).
   \bibitem{CaiPLB2004} R.~G.~Cai, \emph{A Note on thermodynamics of black holes in Lovelock gravity}, Phys.\ Lett.\ B {\bf 582}, 237 (2004), [\href{http://arxiv.org/abs/hep-th/0311240}{{\tt arXiv:hep-th/0311240}}].
   \bibitem{Li:2011uk} P.~Li, R.~H.~Yue and D.~C.~Zou, {\it {Thermodynamics of Third Order Lovelock-Born-Infeld Black Holes}}, Commun.\ Theor.\ Phys.\  {\bf 56}, 845 (2011), [\href{http://arxiv.org/abs/1110.0064}{{\tt arXiv:gr-qc/1110.0064}}].
   \bibitem{Cvetic:2016sow} M.~Cvetic, X.~H.~Feng, H.~Lu and C.~N.~Pope, \emph{Rotating Solutions in Critical Lovelock Gravities}, Phys.\ Lett.\ B {\bf 765}, 181 (2017), [\href{http://arxiv.org/abs/1609.09136}{{\tt arXiv:hep-th/1609.09136}}].
   \bibitem{HennigarJHEP2017} R.~A.~Hennigar, E.~Tjoa and R.~B.~Mann, \emph{Thermodynamics of hairy black holes in Lovelock gravity}, JHEP {\bf 1702}, 070 (2017), [\href{http://arxiv.org/abs/1612.06852}{{\tt arXiv:hep-th/1612.06852}}].
   \bibitem{HennigarPRL2017} R. A. Hennigar, R. B. Mann, E. Tjoa, \emph{Superfluid Black Holes}, Phys.\ Rev.\ Lett.\  {\bf 118}, no. 2, 021301 (2017), [\href{http://arxiv.org/abs/1609.02564}{{\tt arXiv:hep-th/1609.02564}}].
   \bibitem{Planck2018} N.~Aghanim {\it et al.} [Planck Collaboration], \emph{Planck 2018 results. VI. Cosmological parameters}, [\href{http://arxiv.org/abs/1807.06209}{{\tt arXiv:astro-ph/1807.06209}}].
   \bibitem{KiselevCQG2003} V.~V.~Kiselev, \emph{Quintessence and black holes}, Class.\ Quant.\ Grav.\  {\bf 20}, 1187 (2003), [\href{http://arxiv.org/abs/gr-qc/0210040}{{\tt arXiv:gr-qc/0210040}}].
   \bibitem{MaCPL2007} C.~R.~Ma, Y.~X.~Gui and F.~J.~Wang, \emph{Quintessence contribution to a Schwarzschild black hole entropy}, Chin.\ Phys.\ Lett.\  {\bf 24}, 3286 (2007).
   \bibitem{FernandoGRG2012} S.~Fernando, \emph{Schwarzschild black hole surrounded by quintessence: Null geodesics}, Gen.\ Rel.\ Grav.\  {\bf 44}, 1857 (2012), [\href{http://arxiv.org/abs/1202.1502}{{\tt arXiv:gr-qc/1202.1502}}].
   \bibitem{FengPLA2014} Z.~Feng, L.~Zhang and X.~Zu, \emph{The remnants in Reissner-Nordstr$\ddot{o}$m-de Sitter quintessence black hole}, Mod.\ Phys.\ Lett.\ A {\bf 29}, no. 26, 1450123 (2014).
   \bibitem{MalakolkalamiASS2015} B. Malakolkalami and K. Ghaderi, \emph{Schwarzschild-anti de Sitter black hole with quintessence}, Astrophys. Space Sci. {\bf 357}, 112 (2015).
   \bibitem{HussainGRG2015} I. Hussain and S. Ali, \emph{Effect of quintessence on the energy of the Reissner-Nordstr$\ddot{o}$m black hole}, Gen. Rel. Grav. {\bf{47}}, 34 (2015), [\href{http://arxiv.org/abs/1408.3111}{{\tt arXiv:gr-qc/1408.3111}}].
   \bibitem{Ghosh:2016ddh} S.~G.~Ghosh, M.~Amir and S.~D.~Maharaj, \emph{Quintessence background for 5D Einstein-Gauss-Bonnet black holes}, Eur.\ Phys.\ J.\ C {\bf 77}, no. 8, 530 (2017), [\href{http://arxiv.org/abs/1611.02936}{{\tt arXiv:gr-qc/1611.02936}}].
   \bibitem{Ghosh:2017cuq} S. G. Ghosh, S. D. Maharaj, D. Baboolal and T. H. Lee, {\it {Lovelock black holes surrounded by quintessence}}, Eur.\ Phys.\ J.\ C {\bf 78}, no. 2, 90 (2018), [\href{http://arxiv.org/abs/1708.03884}{{\tt arXiv:gr-qc/1708.03884}}].
   \bibitem{Toledo:2018pfy} J.~de M.Toledo and V.~B.~Bezerra, {\it {Black holes with cloud of strings and quintessence in Lovelock gravity}}, Eur.\ Phys.\ J.\ C {\bf 78}, no. 7, 534 (2018).
   \bibitem{Kamenshchik:2001cp} A.~Y.~Kamenshchik, U.~Moschella and V.~Pasquier, \emph{An Alternative to quintessence}, Phys.\ Lett.\ B {\bf 511}, 265 (2001), [\href{https://arxiv.org/abs/gr-qc/0103004}{{\tt arXiv:gr-qc/0103004}}].
   \bibitem{Bilic:2001cg} N.~Bilic, G.~B.~Tupper and R.~D.~Viollier, {\it {Unification of dark matter and dark energy: The Inhomogeneous Chaplygin gas}}, Phys.\ Lett.\ B {\bf 535}, 17 (2002), [\href{https://arxiv.org/abs/astro-ph/0111325}{{\tt arXiv:astro-ph/0111325}}].
   \bibitem{Bento:2002ps} M.~C.~Bento, O.~Bertolami and A.~A.~Sen, {\it{Generalized Chaplygin gas, accelerated expansion and dark energy-matter unification}}, Phys.\ Rev.\ D {\bf 66}, 043507 (2002), [\href{https://arxiv.org/abs/gr-qc/0202064}{{\tt arXiv:gr-qc/0202064}}].
   \bibitem{Carturan:2002si} D.~Carturan and F.~Finelli, \emph{Cosmological effects of a class of fluid dark energy models}, Phys.\ Rev.\ D {\bf 68}, 103501 (2003), [\href{https://arxiv.org/abs/astro-ph/0211626}{{\tt arXiv:astro-ph/0211626}}].
   \bibitem{Amendola:2003bz} L.~Amendola, F.~Finelli, C.~Burigana and D.~Carturan, \emph{WMAP and the generalized Chaplygin gas}, JCAP {\bf 0307}, 005 (2003), [\href{https://arxiv.org/abs/astro-ph/0304325}{{\tt arXiv:astro-ph/0304325}}].
   \bibitem{Bean:2003ae} R.~Bean and O.~Dore, \emph{Are Chaplygin gases serious contenders to the dark energy throne?}, Phys.\ Rev.\ D {\bf 68}, 023515 (2003), [\href{https://arxiv.org/abs/astro-ph/0301308}{{\tt arXiv:astro-ph/0301308}}].
   \bibitem{Torii:2005xu} T.~Torii and H.~Maeda, \emph{Spacetime structure of static solutions in Gauss-Bonnet gravity: Neutral case}, Phys.\ Rev.\ D {\bf 71}, 124002 (2005), [\href{https://arxiv.org/abs/hep-th/0504127}{{\tt arXiv:hep-th/0504127}}].
   \bibitem{Torii:2005nh} T.~Torii and H.~Maeda, \emph{Spacetime structure of static solutions in Gauss-Bonnet gravity: Charged case}, Phys.\ Rev.\ D {\bf 72}, 064007 (2005), [\href{https://arxiv.org/abs/hep-th/0504141}{{\tt arXiv:hep-th/0504141}}].


\end{thebibliography}
\end{document}